\documentclass[preprint,sort&compress]{elsarticle}
\usepackage[T1]{fontenc}
\usepackage{rotating}
\usepackage{amsmath}
\usepackage{amsfonts}
\usepackage{amssymb}
\usepackage{geometry}
\usepackage{graphicx}
\usepackage{caption}
\usepackage{subcaption}
\usepackage{bbm}
\usepackage{epsfig}
\usepackage{pst-grad}
\usepackage{epstopdf}
\usepackage{overpic}
\usepackage{multirow}
\usepackage{tikz}
\usetikzlibrary{calc}
\usepackage{float}

\tikzset{
    >=stealth
}
\journal{Magnetism and Magnetic Materials}

\begin{document}
\begin{frontmatter}

\title{A Device model for xMR Sensors based on the Stoner-Wohlfarth model}

\author[cdlabor]{Florian Bruckner \corref{cor1}}
\ead{florian.bruckner@tuwien.ac.at}
\author[cdlabor]{Bernhard Bergmair}
\author[cdlabor]{Hubert Brueckl}
\author[cdlabor]{Pietro Palmesi}
\author[cdlabor]{Anton Buder}
\author[infineon]{Armin Satz}
\author[cdlabor]{Dieter Suess}

\cortext[cor1]{Corresponding author}

\address[cdlabor] {Christian Doppler Laboratory for Advanced Magnetic Sensing and Materials, Institute of Solid State Physics, Vienna University of Technology, Wiedner Hauptstrasse 8-10, 1040 Vienna, Austria}
\address[infineon] {Infineon Technologies Austria AG, Siemensstrasse 2, 9500 Villach, Austria}

\begin{abstract}
The Stoner-Wohlfarth model provides an efficient analytical model to describe the behavior of magnetic layers within xMR sensors. Combined with a proper description of magneto-resistivity an efficient device model can be derived, which is necessary for an optimal electric circuit design. Parameters of the model are determined by global optimization of an application specific cost function which contains measured resistances for different applied fields. Several application cases are examined and used for validation of the device model. Furthermore the applicability of the SW model is verified by comparison with micromagnetic energy minimization results.
\end{abstract}


\end{frontmatter}

\section{Introduction}
\label{sec:Introduction}
Magnetic sensors offer the possibility of contact-less sensing and are used in a wide field of applications \cite{daughton_gmr_1999}. Read-head sensors for magnetic storage devices, magnetic bio-sensors and position-, speed- and binary-state sensors used in automotive- and industrial-applications are only a few examples. Compared with the well established hall sensors xMR sensors offer higher sensitivity and thus allow a more flexible use of the sensors. Furthermore it is possible to combine the sensing element and signal conditioning electronics into one robust, monolithic design \cite{kapser_integrated_2007, oslon_integrated_2005}.

For an optimal design of the signal condition circuit a detailed description of the sensing element is of utmost importance. While micromagnetic simulations with well-established tools \cite{fidler_micromagnetic_2000} provide an accurate description of the magnetic state, those simulations are very time-consuming and it is difficult to directly interface them with electric circuit simulations. Device models based on electrical equivalent circuits \cite{das_generalized_1999, lee_advanced_2005} are well suited for integration in various hardware simulation environments like SPICE, VHDL-AMS or VERILOG-A. However using a modeling approach based on simplified physical assumptions is advantageous because it provides insight into the underlying physics and thus leads to more extensible models. In Ref. \cite{kammerer_compact_2004} a device model for a magnetic-tunnel-junction(MTJ) has been presented, which iteratively solves a Stoner-Wohlfarth model for a quasi-static description of the magnetic layers. Using the Landau-Lifshitz-Gilbert equation extends the model to dynamic magnetization processes \cite{kammerer_compact_2010, jander_dynamic_2008}.

Within this work an analytical solution of the Stoner-Wohlfarth model is used which provides an efficient and accurate description of the magnetic properties of the xMR sensor. In practice the subjective selection of optimal parameters for different models complicates the validation of model features. A generic and flexible method, based on the global optimization of an application specific cost function, is proposed to automatically determine reproducible optimal parameters for the model. Furthermore the model is applied to different test cases and compared with measurement results. Additionally micromagnetic simulations can be used for validation of the device model without the need for time-consuming measurements.

The structure of the paper is as follows. Section \ref{sec:Device model} summarizes electric and magnetic properties of xMR sensors which will be described by the device model and explains in detail how the underlying Stoner-Wohlfarth model can be solved analytically. The solution of the global optimization problem introduced in Sec. \ref{sec:Optimizations} determines the optimal device parameters for a certain application case. Finally in Sec. \ref{sec:Measurements} the device model is validated by comparison with measurements for several test cases.

\section{Device model}
\label{sec:Device model}
xMR sensors consist of several thin magnetic layers which change their magnetization depending on the applied field, which in turn influences the resistance of the sensor. Although it is possible to directly measure the magnetic state of a magnetic layer (e.g. using the magneto-optic Kerr effect), it is more convenient to simply measure the total resistance of the device. Therefore it is necessary to model both the magnetic and the electric properties of the device.

\subsection{Electric Properties}
Within this work the following types of magneto-resistance have been taken into account:
\begin{itemize}
\item{\bf Anisotropic magnetoresistance (AMR):}
AMR sensors only consist of a single magnetic layer which changes its resistance according to the angle $\phi_\text{AMR}$ between its magnetization and the applied electrical current. The AMR effect is a uniaxial effect, which can be described by the following relation \cite{campbell_spontaneous_1970}
\begin{align}
dR_\text{AMR} = C_\text{AMR} \cos^2({\phi_\text{AMR}})
\end{align}

\item{\bf Giant magnetoresistance (GMR), Tunnel magnetoresistance (TMR):}
Both GMR \cite{baibich_giant_1988} and TMR \cite{julliere_tunneling_1975} sensors contain at least two magnetic layers. Within this work we are focusing on so called spin-valve sensors \cite{freitas_magnetoresistive_2007} which consist of a soft-magnetic free-layer and at least one pinned-layer. The magnetization of the free-layer is strongly influenced by the external field, whereas the pinned-layer determines a reference direction independent of the external field. The magnetization of the pinned-layer is fixed via exchange-coupling with an anti-ferromagnetic layer, which reduces the effective saturation magnetization $M_s$ and thus increases the magnetic anisotropy field $H_k$. Pinning strength can be further improved by using two anti-ferromagnetically coupled pinned-layers. The GMR as well as the TMR effect show unidirectional behavior
\begin{align} \label{eqn:gmr_effect}
dR_\text{GMR} = C_\text{GMR} \cos({\phi_\text{GMR}}) & & 
dR_\text{TMR} = C_\text{TMR} \cos({\phi_\text{TMR}})
\end{align}
, where $\phi_\text{GMR}$ and $\phi_\text{TMR}$ is the relative angle between the magnetizations within free- and pinned-layer. Despite the more complex structure of these sensors they a preferred for applications, because the provide larger sensitivities \cite{daughton_gmr_1999}.
\end{itemize}

\subsection{Magnetic Properties}
The lateral dimensions of the xMR sensors under consideration vary from $<1\mu m$ to $100\mu m$. The Stoner-Wohlfarth(SW) model provides an analytical model to describe the magnetic layers as long as they behave like single-domain particles. This is the case especially for small sensors, but even for larger ones the SW model gives good approximations in many application cases. 

\begin{figure}[h!]
\centering
\begin{overpic}[width=0.15\columnwidth]{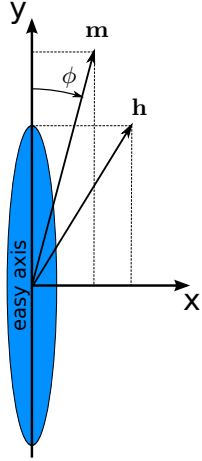}
	\put(12,82){\small $\phi$}
%	\put(20,68){\small $\theta$}
	\put(17,92){\small $\mathbf{m}$}
	\put(27,75){\small $\mathbf{h}$}
\end{overpic}

\caption{\small Visualization of a Stoner-Wohlfarth particle with an uniaxial anisotropy along the y-axis (easy-axis). The applied field $\mathbf{h}$ leads to a magnetization $\mathbf{m}$ with a fixed modulus of $M_s$.}
\label{fig:sw_model}
\end{figure}

\begin{figure}[H]
\begin{center}
\begin{overpic}[width=0.50\columnwidth]{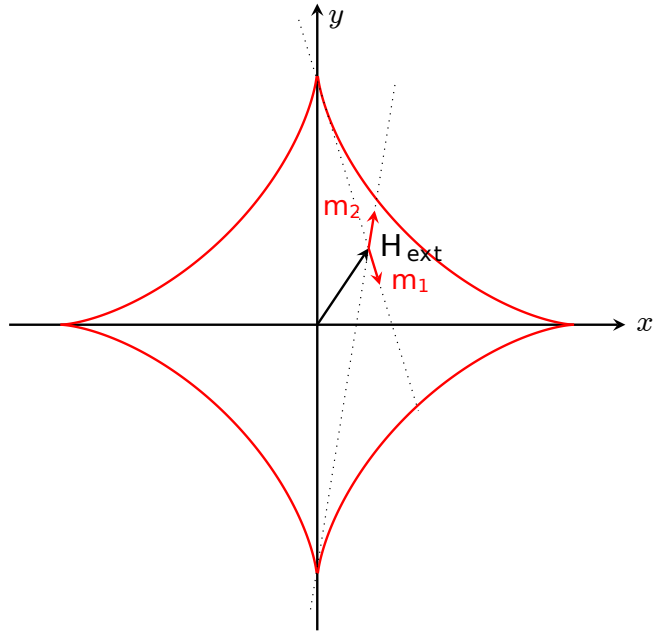}
\end{overpic}
%\begin{tikzpicture}[scale=3,smooth,domain=-1.2:1.2];
%% Draw axes
%    \draw [->,thick] (-1.2,0.0) coordinate (xaxis) |- (1.2,0.0) node [right] {$x$};
%    \draw [->,thick] (0.0,-1.2) coordinate (yaxis) |- (0.0,1.2) node [right] {$y$};
%
%% Draw Asteroid
%\draw[color=red, thick, samples=500] plot[id=sin, domain=-1:1] function{(1.-x**(2./3.))**(3./2.))};
%\draw[color=red, thick, samples=500] plot[id=sin, domain=-1:1] function{-(1.-x**(2./3.))**(3./2.))};
%\draw[color=red, thick, samples=500] plot[id=sin, domain=-1:1] function{(1.-(-x)**(2./3.))**(3./2.))};
%\draw[color=red, thick, samples=500] plot[id=sin, domain=-1:1] function{-(1.-(-x)**(2./3.))**(3./2.))};
%
%% Draw field and tangents
%\coordinate (P) at (0.20, 0.30);
%\coordinate (T1) at (0.044, 0.814);
%\coordinate (T2) at (0.000, -0.953);
%
%\draw [->, thick](0.0,0.0) -- (P) node [right] {$\mathbf{H_{ext}}$};
%%\draw[dotted] (yaxis |- P) -| (xaxis -| P);
%\draw[dotted, shorten >= -2.0cm, shorten <= -1.2cm] (T1) -- (P);
%\draw[dotted, shorten >= -2.0cm, shorten <= -0.5cm] (T2) -- (P);
%
%% Draw Magnetizations 
%\draw [->, color=red, thick] (P) -> ($(P)!0.15cm!180:(T1)$) node [right] {\small{$\mathbf{m_1}$}};
%\draw [->, color=red, thick] (P) -> ($(P)!0.15cm!180:(T2)$) node [left] {\small{$\mathbf{m_2}$}};
%
%\end{tikzpicture}
\caption{Stoner-Wohlfarth asteroid for uniaxial anisotropy along $y$ axis. Inside of the asteroid there exist two stable solutions which can be constructed by tangents on the asteroid which pass through the tip of the applied field vector $H_{ext}$. Outside of the asteroid there exists only a single stable solution.}
\label{fig:sw_asteroid}
\end{center}
\end{figure}

The SW model assumes a single-domain inside of a layer which allows to describe the magnetization state as a single unit magnetization vector $\mathbf{m} = \frac{\mathbf{M}}{M_s}$. Only an external field and (uniaxial) anisotropy are considered. Due to the symmetry of the problem the magnetization lies within the plane spanned by the external field and the anisotropy axis, which reduces the problem to a 2D problem. The reduced total energy density reads as 
\begin{align}
\eta = \underbrace{\frac{1}{2} \sin^2(\phi)}_\text{Anisotropy energy} \; \underbrace{-h_x \sin(\phi) -h_y \cos(\phi)}_\text{Zeeman energy} 
\end{align}
, with $\eta = \frac{E}{2 K V}$, $\mathbf{h} = \frac{\mathbf{H}}{H_k}$ and $H_k = \frac{2 K}{\mu_0 M_s}$. $V$ is the volume of the layer, $K$ the uniaxial anisotropy constant, $\mu_0$ the permeability of the free space, and $\phi$ is the angle between anisotropy easy-axis which is assumed along the $y$-axis (see Fig. \ref{fig:sw_model}). The stable magnetization states for an applied field can be calculated analytically, by a transformation to Cartesian coordinates as proposed in Ref. \cite{wood_exact_2009}. The stability conditions requires that the first derivative of the total energy vanishes.
 
\begin{align}
\frac{\partial \eta}{\partial \phi} = 0 \quad \Rightarrow \quad \frac{h_x}{m_x} - \frac{h_y}{m_y} = 1
\end{align}
Considering that $m_x^2 + m_y^2 = 1$ since $\mathbf{m}$ is the reduced magnetization allows to eliminate either $m_x$ or $m_y$ and leads to one of the following 4th order equations:
\begin{subequations} \label{eqn:sw_4order}
\begin{align}
m_x^4 - 2 H_x \; m_x^3 + (h^2-1) \; m_x^2 + 2 h_x \; m_x - h_x^2 &= 0 \\
m_y^4 + 2 H_y \; m_y^3 + (h^2-1) \; m_y^2 - 2 h_y \; m_y - h_y^2 &= 0
\end{align}
\end{subequations}
In general a 4th order equation always has 4 solutions, however not all of them need to be real. Considering $\frac{\partial^2 \eta}{\partial \phi^2} = 0$ allows to define a critical curve which separates the inner region where all four solutions are real from outer region where only 2 real solutions exist. This critical curve is also called Stoner-Wohlfarth asteroid (see Fig. \ref{fig:sw_asteroid}) and written in Cartesian coordinates it looks like
\begin{align}
h_x^{2/3} + h_y^{2/3} = 1
\end{align}

Solving equation \eqref{eqn:sw_4order} is usually done by reducing the quartic to a subsidiary cubic equation. A real root of the cubic equation is then used to factorize the quartic to quadratic equations which can easily be solved. Different approaches to reduce the quartic equations lead to numerical instabilities for different input coefficients. Combination of Ferrari's, Descartes', Neumark's, or Yacoub \& Fraidenraich's algorithms depending on the actual coefficients \cite{paeth_graphics_1995} produces stable results for arbitrary applied fields.

\section{Parameter Optimizations}
\label{sec:Optimizations}
Identifying all parameters of a device model can be a complicated task since only limited measurement capabilities are available. Additionally for a real sensor the assumptions of the idealized model may be violated to a certain extent. An alternative and very comfortable way of selecting the necessary parameters is by determining a cost-function and solving an optimization problem. Independent of the type and extent of the measurement data this method provides an optimal device model. The choice of the cost function furthermore allows to focus the optimization on certain features which are most important for the current application.

Finding the optimal device model is a multidimensional global optimization problem (GLP). In contrast to local optimization problems most deterministic algorithms for global optimization scale very badly. Different heuristic methods have been proposed to overcome the performance problem. Within this work different GLP algorithms from the OpenOpt \cite{kroshko_dmitrey_openopt:_2007} suite have been applied to the parameter optimization problem and their performance and accuracy have been compared. Due to the unified interface of OpenOpt switching between the provided optimization methods proves to be simple.  

\begin{figure}[h!]
\centering
\begin{overpic}[width=0.9\columnwidth]{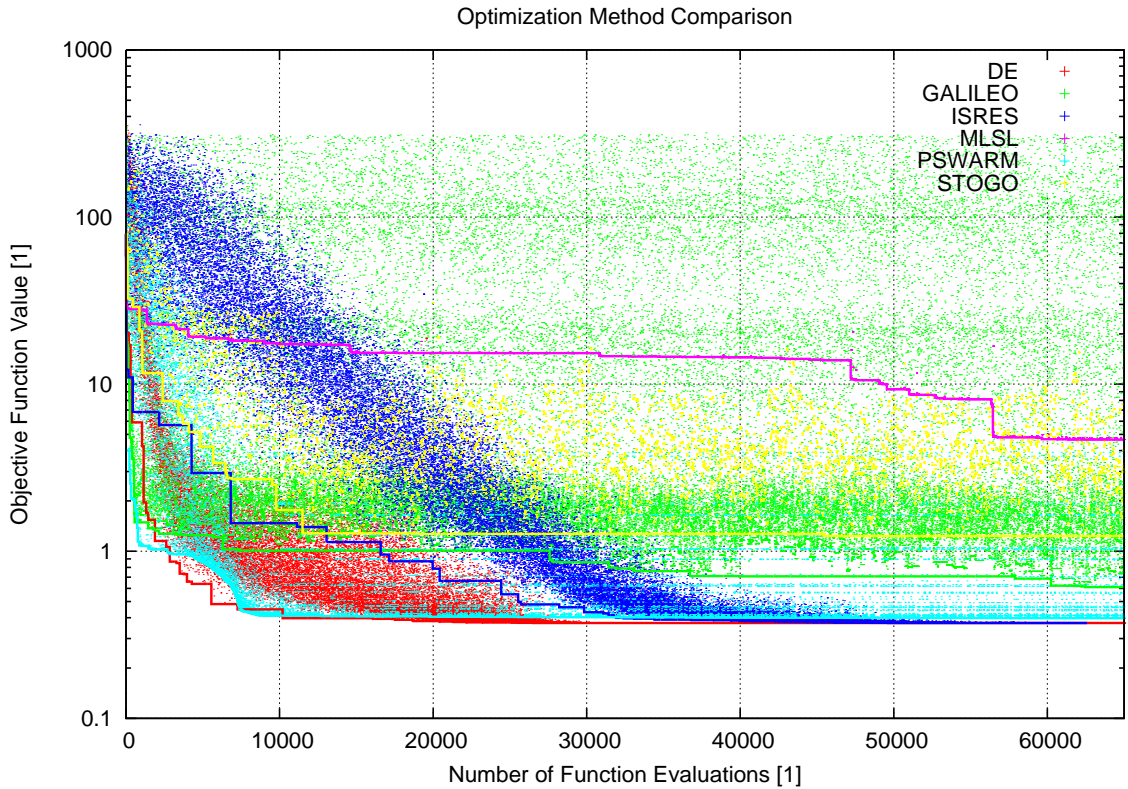}
\end{overpic}
\caption{\small Performance comparison of different global optimization methods for the optimization of a device model using 8 design variables (see Sec. \ref{sec:Measurements}). Each objective function evaluation as well as the actual minimum values are plotted. After 60000 function evaluations DE and ISRES seem to be converged whereas the other algorithms perform worse for the current problem. The achieved objective function as well as the corresponding design variables are listed in Tbl. \ref{tbl:optimizer_comparison}. }
\label{fig:optimizer_comparison}
\end{figure}

The objective function $O$ is defined as 
\begin{align}
O = \sum\limits_{i=1}^N w_i \left( x_i - x^m_i \right)^2
\end{align}
, where $N$ is the number of measurements, $w_i$ is the weight of the $i$-th measurement, $x_i$ is the simulated result and $x^m_i$ is the measured result. Within this work only equally weighted measurements $w_i=1$ were used. The resistance of the xMR element was measured as a function of the applied field strength for various field angles as described in Sec. \ref{sec:Measurements}. A comparison of different GLP methods can be found in Fig. \ref{fig:optimizer_comparison} as well as in Tbl. \ref{tbl:optimizer_comparison}. The DE algorithm proved most efficient and reliable and will be used for all optimizations done in the following sections.

\begin{table}[h!]
\centering
\begin{tabular}{|c||c|c|c|c|c|c|c|c|c|}
\hline
\multirow{3}{*}{Method} & Objective function & \multicolumn{4}{|c|}{free-layer}                                             & \multicolumn{2}{|c|}{pinned-layer} & \multicolumn{2}{|c|}{transfer-function} \\
                        & $O$                & $\mu_0 H_k$ & $\phi$     & $\mu_0 H^\text{bias}_x$ & $\mu_0 H^\text{bias}_y$ & $\mu_0 H_k$ & $\phi$               & $R_0$       & $dR_\text{GMR}$ \\
                        & $[1]$              & $[mT]$      & $[^\circ]$ & $[mT]$                  & $[mT]$                  & $[mT]$      & $[^\circ]$           & $[k\Omega]$ & $[\Omega]$ \\
\hline
\hline
de       & 0.372 & 10.00 & 90.0 &  2.00 & -1.19 & 405.81 &  -1.61 & 18.758 & 897.91 \\
isres    & 0.372 & 10.00 & 90.0 &  1.99 & -1.18 & 405.57 &  -1.60 & 18.758 & 897.81 \\
pswarm   & 0.393 & 10.54 & 90.0 &  0.15 & -1.70 & 561.82 &  -1.56 & 18.756 & 898.11 \\
mlsl     & 0.745 &  8.85 & 90.0 &  2.00 &  1.55 & 323.17 &  -1.53 & 18.792 & 892.44 \\
heeds    & 0.911 & 10.12 & 88.2 & -1.88 &  1.24 & 555.00 &  -1.80 & 18.800 & 900.82 \\
galileo  & 0.987 &  7.55 & 86.9 &  1.60 & -1.82 & 377.94 &  -3.41 & 18.749 & 900.89 \\
stogo    & 1.191 & 17.02 & 46.4 & -1.43 &  0.42 & 639.18 & -12.08 & 18.794 & 1009.75 \\
\hline
\end{tabular}
\caption{\small Comparison of the optimal solution of different global optimization methods for the optimization of a device model using 8 design variables. It can be seen that for the algorithms which converged to the minimal objective function (DE, ISRES; see Fig. \ref{fig:optimizer_comparison}) also the optimal parameters match very well. }
\label{tbl:optimizer_comparison}
\end{table}

\section{Measurements}
\label{sec:Measurements}

\subsection{Hysteresis measurements}
One key feature of the SW model is the representation of a hysteresis. For the characterization of an xMR sensor hysteresis curves were measured along certain directions. Depending on the field angle the behavior of the sensor changes dramatically. For applied fields orthogonal to the easy-axis the magnetization of a SW particle shows linear behavior, whereas for fields along the easy-axis it shows abrupt switching from one state into the other. 

In order to characterize the direction dependent hysteresis behavior of the sensor a homogeneous field created by a Helmholtz coil setup is applied. After the sensor is saturated applying a field of $50mT$, field amplitude is decreased down to $-50mT$ and again increased up to $+50mT$ using $0.5mT$ steps. Field angles are adjusted in $10^\circ$ steps by rotating the whole sensor against the stationary Helmholtz coil. The measured sensor-resistance is used to find the optimal parameters of a certain device models, by minimizing the squared difference of measured and simulated resistance. 

\begin{figure}[h!]
        \centering
        \begin{subfigure}[b]{0.5\textwidth}
                \includegraphics[width=\textwidth]{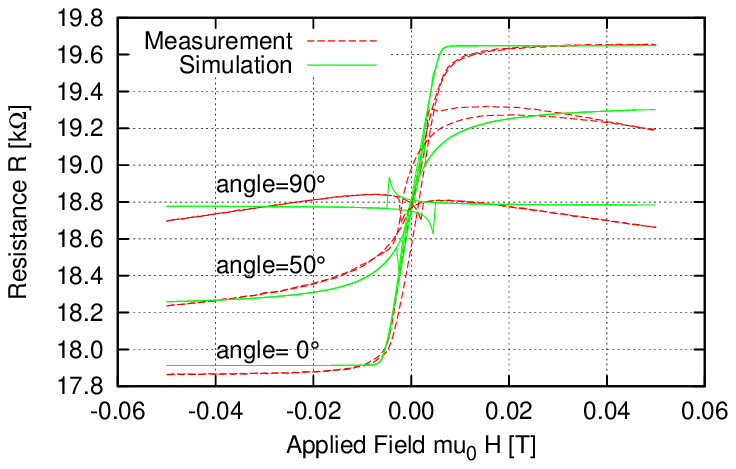}
                \caption{Device model Stage I}
                \label{fig:stageI}
        \end{subfigure}~
        \begin{subfigure}[b]{0.5\textwidth}
                \includegraphics[width=\textwidth]{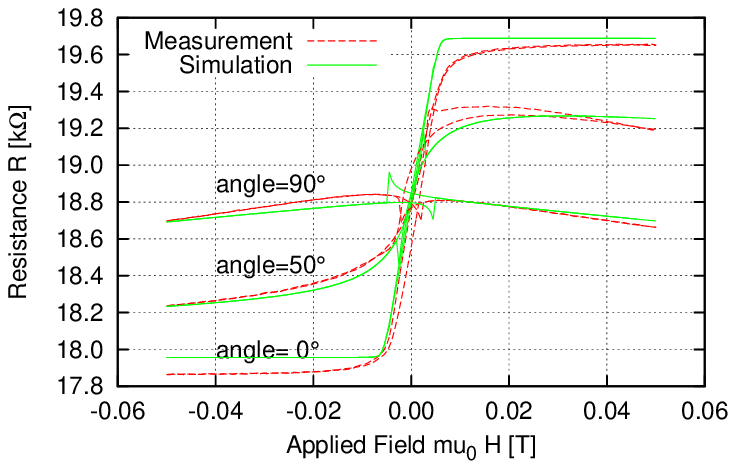}
                \caption{Device model Stage II}
                \label{fig:stageII}
        \end{subfigure}

        \begin{subfigure}[b]{0.5\textwidth}
                \includegraphics[width=\textwidth]{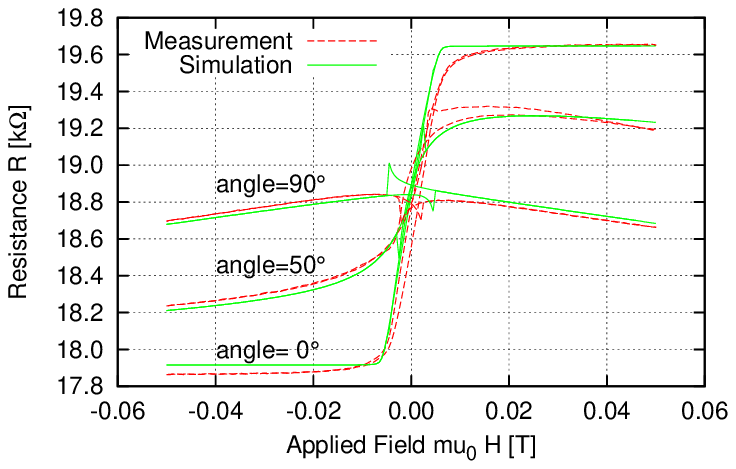}
                \caption{Device model Stage III}
                \label{fig:stageIII}
        \end{subfigure}~
        \begin{subfigure}[b]{0.5\textwidth}
                \includegraphics[width=\textwidth]{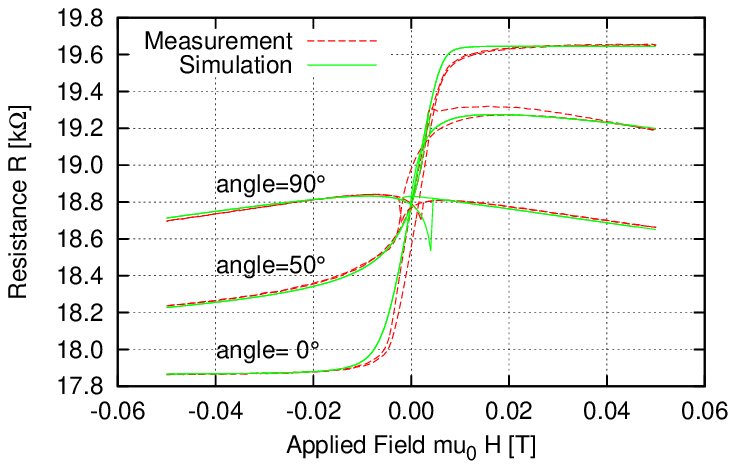}
                \caption{Device model Stage IV}
                \label{fig:stageIV}
        \end{subfigure}
        \caption{Development of the simulated hysteresis for field angles of $0^\circ$, $50^\circ$, and $90^\circ$ at different modeling stages of the device model. (a) Only the magnetic freelayer simulated, (b) magnetic free- and pinned-layer are considered, (c) electrical AMR effect within free-layer is added, (d) a bias-field describing the uncompensated interaction from the pinned-layer onto the free-layer is handled. The optimized parameters as well as the achieved objective functions can be found in Tbl. \ref{tbl:device model_development}. }\label{fig:device model_stages}
\end{figure}

The following list summarizes the progress of the device model and describe which features of the measurement data can be reproduced by the different stages. Measured and simulated hysteresis curves are visualized in Fig. \ref{fig:device model_stages} and the optimal parameters can be found in Tbl. \ref{tbl:device model_development}.

\begin{itemize}
\item{\bf Stage I: } (see Fig. \ref{fig:stageI})
The simplest model of a xMR sensor only considers its magnetic free-layer. The anisotropy field $H_k$ as well as the easy-axis angle $\phi$ are used as design parameters. The pinned-layer is replaced with a fixed reference direction which is necessary to calculate the resistance change according to Eqn. \eqref{eqn:gmr_effect}, with $\phi_\text{GMR} = \phi$. 

Although this model is very simple the qualitative behavior of the hysteresis for different angles can already be reproduced (e.g. the small hysteresis of the $0^\circ$ curve results from a tilting of the freelayer easy-axis). Nevertheless there are still large deviations from the measurement. Especially at large field amplitudes the simulated resistance saturates, whereas the measured data show a non-zero slope. 

\item{\bf Stage II: } (see Fig. \ref{fig:stageII})
The magnetic properties of the pinned layer ($H_k^P$, $\phi^P$) are added to the model. 

Compared with the magnetic freelayer the pinned layer has a much larger anisotropy field, due to the small effective $M_s$ of the synthetic anti-ferromagnet. Therefore it is less resistive on the external field. Thus the sensor output is only slightly modified for small field amplitudes. For large fields however the pinned layer is significantly deflected which explains the non-zero slope of the measured resistance.

\item{\bf Stage III: } (see Fig. \ref{fig:stageIII})
The simulated resistance at large field amplitudes still shows an angle dependent offset. An AMR effect within the free-layer is assumed ($dR_\text{AMR}$) which allows to compensate this offset. 

\item{\bf Stage IV: } (see Fig. \ref{fig:stageIV})
Finally a stationary bias field can be added which allows to describe an interaction from the pinned layer onto the freelayer. Usually the strayfield interaction between the two layers can be compensated by means of a tunable RKKY interaction. Variation of the thickness of the thickness of the conducting layer (GMR) or the tunnel barrier (TMR) allows to control the RKKY interaction strength. Unfortunately the compensation only works in average and it even may be incomplete due to a non-optimal production process. 

A remaining interaction can explain a broken symmetry in the sensor output as it can be observed in the measurement data of the $90^\circ$ hysteresis. Two jumps from lower resistance to higher resistance occur, whereas an unbiased SW model predicts a symmetric response. Adding an uncompensated $x$-bias-field to the SW model allows to reproduce this effect.
\end{itemize}

\begin{table}[h!]
\centering
\begin{tabular}{|c||c|c|c|c|c|c|c|c|c|c|}
\hline
\multirow{3}{*}{Model description} & Objective  & \multicolumn{4}{|c|}{free-layer}                                             & \multicolumn{2}{|c|}{pinned-layer} & \multicolumn{3}{|c|}{transfer-function} \\
                                   & function   & $\mu_0 H_k$ & $\phi$     & $\mu_0 H^\text{bias}_x$ & $\mu_0 H^\text{bias}_y$ & $\mu_0 H^P_k$ & $\phi^P$           & $R_0$       & $dR_\text{GMR}$ & $dR_\text{AMR}$\\
                                   & $O \; [1]$ & $[mT]$      & $[^\circ]$ & $[mT]$                  & $[mT]$                  & $[mT]$        & $[^\circ]$         & $[k\Omega]$ & $[\Omega]$ & $[\Omega]$\\
\hline
\hline
only freelayer   & 3.013 &  5.64 & 88.0 &   -   &   -   &    -   &   -    & 18.781 & 867.37 &   -   \\
with pinned layer & 2.640 &  5.65 & 88.3 &   -   &   -   & 338.78 &   0.00 & 18.822 & 865.83 &   -   \\
with AMR effect  & 2.306 &  5.65 & 88.3 &   -   &   -   & 238.36 &   0.00 & 18.866 & 865.15 & 85.57 \\
with bias field  & 0.537 &  4.51 & 84.1 & -0.48 & -1.73 & 231.78 &  -2.66 & 18.818 & 890.57 & -127.17 \\
\hline
\end{tabular}
\caption{\small Optimal parameters of the device model at different modeling stages. Optimization has been done for all specified values. The corresponding hysteresis curves can be seen in Fig. \ref{fig:device model_stages}. }
\label{tbl:device model_development}
\end{table}

\subsection{Asteroid measurement}
\begin{figure}[h!]
\centering
\begin{overpic}[width=0.5\columnwidth]{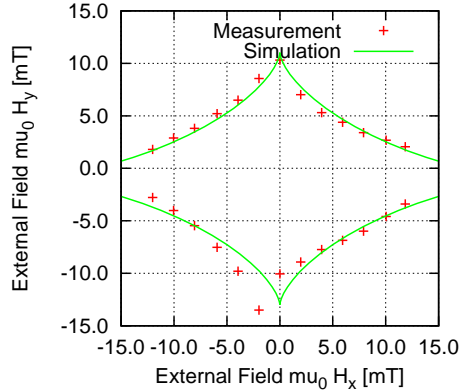}
\end{overpic}

\caption{\small Measurement of the critical field of a GMR sensor for different field directions. The lateral size of element is $390\text{nm} \times 180\text{nm}$ which is small enough to obtain single domain behavior and makes it well suited for a description via the SW model. The saturation field along the hard-axis (24mT) is approximately twice as high as the switching field along the easy-axis (12mT). The measurement data are fitted using a scaled SW model, where only the $y$-component of the applied field is scaled.}
\label{fig:measurement_asteroid}
\end{figure}

Another interesting property of xMR sensors is the switching between the different magnetic states of the free-layer, as described by the SW model. Especially the critical fields at which the switching occurs may be of interest for different applications. Measurements of the critical field along multiple directions show that in contrast to the SW model the critical field along the hard- and easy-axis differ (see Fig. \ref{fig:measurement_asteroid}). This effect can simply be incorporated into the standard SW model by applying a scaled field $\mathbf{h}^*$ to a standard SW model, where only the $y$ component is reduced. 
\begin{align}
h_x^* = h_x & & h_y^* = C \; h_y
\end{align}

A more sophisticated approach directly allows to define a modified critical curve and derive the corresponding energy density for the definition of a generalized SW-model \cite{cimpoesu_generalized_2013}. Using this approach allows refined modifications of the angular dependence of the switching field. The drawback of the method is that it no longer provides an analytical solution for the stable magnetizations.

\subsection{Non Stoner-Wohlfarth behavior}
During the development of the device model some cases could be observed where the sensor output significantly deviates from a simple SW behavior. As a simple test-case a periodic field along the $x$-axis combined with a $45^\circ$ offset field $\mathbf{H_\text{off}}$ is applied
\begin{align} \label{eqn:diagonal_offset_field}
\mathbf{H_\text{ext}} = \begin{pmatrix} H_x \, \sin(\omega t) \\ 0 \end{pmatrix} + \begin{pmatrix} H_\text{off} \\ H_\text{off} \end{pmatrix}
\end{align}
A fixed amplitude of the periodic signal $\mu_0 H_x = 1mT$ is assumed, whereas the amplitude of the offset field is varied. For small offset amplitudes the applied field lies completely inside of the Stoner-Wohlfarth asteroid. Thus according to the SW model the resulting output should depend on the actual state of the sensor. Initializing the sensor with a large field in either $+y$ (UP) or $-y$ (DOWN) direction and comparing the resulting amplitudes of the sensor resistance for different offsets thus allows validation of the SW model. 

\begin{figure}[h!]
\centering
	\includegraphics[width=0.5\textwidth]{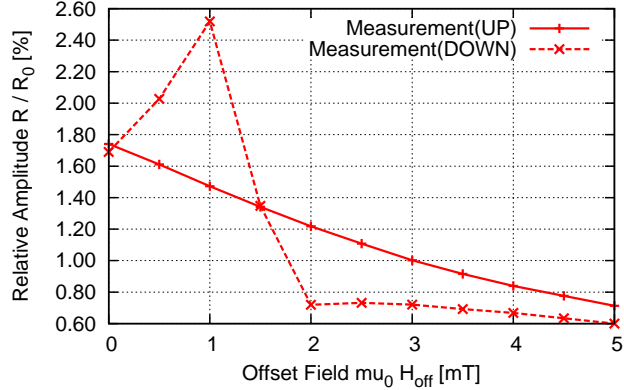}
	\caption{Measured amplitudes of the GMR sensor output signal with an applied oscillating field and a variable offset field according to Eqn. \eqref{eqn:diagonal_offset_field}. The measured resistance amplitudes $R$ relative to $R_0$ are shown for the sensor initialized in either UP($+y$) or DOWN($-y$) state. The crossover of the two branches indicates that inhomogeneous switching occurs. If the applied offset field is further increased the sensor finally saturates and both branches are expected to merge.} 
	\label{fig:measB}
\end{figure}

Measurement results presented in Fig. \ref{fig:measB} show SW behavior for small offset fields. When the offset field is large enough switching occurs, but in contrast to the SW model the sensor state does not completely switch into the opposite state. For even higher offset fields the sensor finally switches completely and again shows SW behavior where both branches coincide.

\begin{figure}[h!]
\centering
	\includegraphics[width=0.5\textwidth]{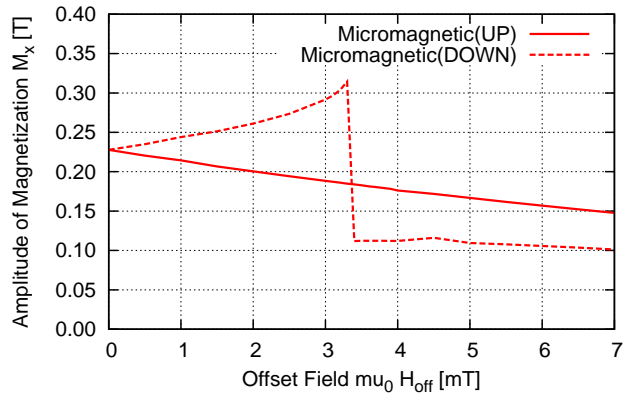}
	\caption{Simulated magnetization amplitude of Sensor A with an applied oscillating field and a variable offset field according to Eqn. \eqref{eqn:diagonal_offset_field}. The average magnetization amplitudes $M_x$ are presented for the sensor initialized in either UP($+y$) or DOWN($-y$) state. The pinned layer is assumed to be aligned along the $-x$ axis, which makes $M_x$ proportional to the resistance $R$ of the sensor. As with the measurement a crossover of both branches occurs. Inspecting the micromagnetic state of the sensor (see Fig. \ref{fig:micromagnum_states}) reveals the nature of the incomplete switching process. A small region next to the boundary remains unswitched and thus reduces the effective sensor width. This leads to an increased shape anisotropy and in turn corresponds with a decreased magnetization amplitude.}
	\label{fig:simB}
\end{figure}

\begin{figure}[h!]
\centering
	\includegraphics[width=0.5\textwidth]{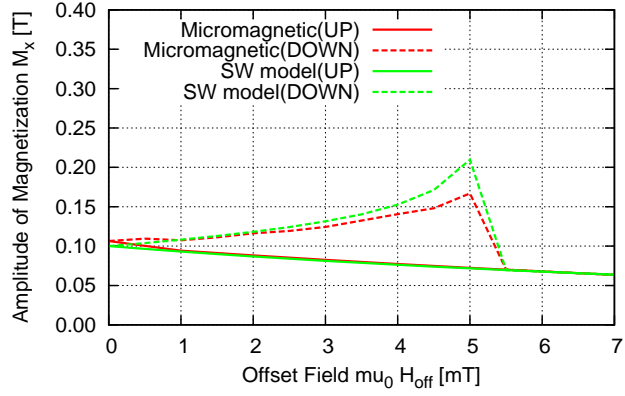}
	\caption{Simulated magnetization amplitude of the reduced-size Sensor B with an applied oscillating field and a variable offset field according to Eqn. \eqref{eqn:diagonal_offset_field}. The average magnetization amplitudes $M_x$ are presented for the sensor initialized in either UP($+y$) or DOWN($-y$) state. The reduction of the sensor size leads to SW behavior independent of the offset field due to a more uniform switching process. The corresponding micromagnetic states of sensor B are visualized in Fig. \ref{fig:micromagnum_states}.}
	\label{fig:simA}
\end{figure}

\begin{figure}[h!]
\centering
\begin{overpic}[width=0.70\columnwidth]{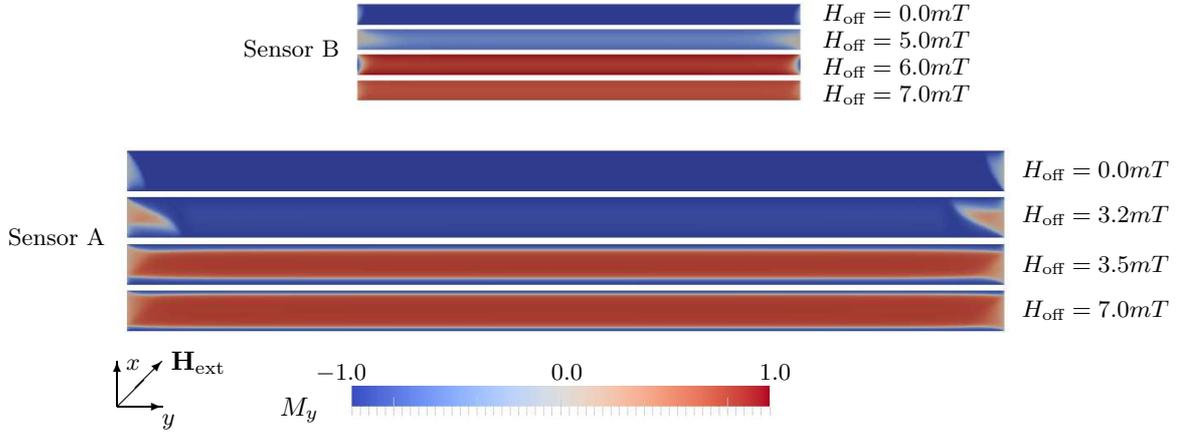}
	\put(14,41){\small Sensor B}
	\put(-12,20){\small Sensor A}
	\put(22,5){\small $-1.0$}
	\put(48,5){\small $0.0$}
	\put(71,5){\small $1.0$}
	\put(18,1){$M_y$}
	\put(0,2){\vector(1,0){5}}
	\put(0,2){\vector(0,1){5}}
	\put(0,2){\vector(1,1){5}}
	\put(6,6){$\mathbf{H_\text{ext}}$}
	\put(1,6){\small $x$}
	\put(5,0){\small $y$}

	\put(78,45.0){\small $H_\text{off} = 0.0mT$}
	\put(78,42.0){\small $H_\text{off} = 5.0mT$}
	\put(78,39.0){\small $H_\text{off} = 6.0mT$}
	\put(78,36.0){\small $H_\text{off} = 7.0mT$}

	\put(100,27.5){\small $H_\text{off} = 0.0mT$}
	\put(100,22.5){\small $H_\text{off} = 3.2mT$}
	\put(100,17.0){\small $H_\text{off} = 3.5mT$}
	\put(100,12.0){\small $H_\text{off} = 7.0mT$}
\end{overpic}

\caption{\small Micromagnetic simulation of Sensor A and B for different applied offset fields $H_\text{off}$ and an additional dynamic $x$-field of $1mT$. The visualization shows the magnetic state after one period of the oscillating field. Both sensors were initially magnetized in $-y$ direction. The magnetization of Sensor B switches completely into $+y$ direction, whereas within Sensor A only an interior region where the demagnetization field is high enough reverses its state. The remaining boundary domain is stabilized by the strayfield of the switched interior region. Even larger external fields are required to trigger a complete switching process.}
\label{fig:micromagnum_states}
\end{figure}

The partly switching process can be explained by micromagnetic simulations which allow to gain an insight into the detailed magnetization states that occur within the magnetic layers. Using an efficient energy minimization method \cite{abert_efficient_2014} to determine stationary magnetic states provides a large performance benefit compared with time integration methods. The magnetization of the magnetic free-layer is simulated for different dimensions and material properties listed in Tbl. \ref{tbl:micromagnetic_properties}. Simulation results using the original sensor dimensions (Sensor A) show good qualitative agreement with the measurement data (see Fig. \ref{fig:simB}). Additionally it can be shown that reducing the sensor dimension (Sensor B) suppresses the formation of the boundary domains. Thus partly switching can be avoided and pure SW behavior is observed (see Fig. \ref{fig:simA}). The magnetization state for various offset fields for both simulations is visualized in Fig. \ref{fig:micromagnum_states}.

\begin{table}[h!]
\centering
\begin{tabular}{|c||c|c|c|c|c|c|c|c|c|c|}
\hline
& Dimension & Saturation & Exchange \\
& $L_x \times L_y \times L_z$ & Polarization $J_s$ & Constant $A$ \\
& [$\text{nm} \times \text{nm} \times \text{nm}$] & [T] & [J/m] \\
\hline
\hline
Simulation Sensor A & $700 \times 15000 \times 5$ & 1.93 & 1.53e-11 \\
Simulation Sensor B & $350 \times  7500 \times 5$ & 0.80 & 2.00e-11 \\
\hline
\end{tabular}
\caption{\small Geometrical and material properties of the micromagnetic models of sensor A and B. The smaller size of sensor B is compensated with adjusted magnetic properties in order to end up with a comparable anisotropy field $H_k$.}
\label{tbl:micromagnetic_properties}
\end{table}

\section{Conclusion}
\label{sec:Conclusion}
An efficient device model consisting of multiple SW particles was presented and applied to different application cases. Different measurements have been used for validation of the model and the missing model parameters are determined by solving a global optimization problem. Hysteresis as well as switching behavior of different sensors have been compared and could be reproduced well. Detailed micromagnetic simulations provide an explanation for deviations from SW behavior in large sensor structures. Depending on size and geometry of the magnetic layer multi-domain states may occur, which violate a basic assumption of the SW model.

\section*{Acknowledgment}
\label{sec:Acknowledgement}
The authors would like to thank the Austrian Federal Ministry of Economy, Family and Youth and the National Foundation for Research, Technology and Development for the financial support.

\bibliographystyle{elsarticle-num}
\bibliography{paper}{}

\end{document}